\documentstyle[preprint,aps,epsfig]{revtex}
\begin{document}
\tightenlines
\author{Chueng-Ryong Ji and Yuriy Mishchenko}
\title{Time-to-space conversion in quantum field theory of flavor mixing}
\date{\today}
\maketitle

\begin{abstract}
We consider the problem of time-to-space conversion in quantum field theory of 
flavor mixing using a generalization of the wave-packet
method in quantum mechanics. 
We work entirely within the canonical formalism of
creation and annihilation operators that allows us, unlike the usual
wave-packet formulation, to include the nontrivial effect due to flavor
condensation in the vacuum.
\end{abstract}

\section{Introduction}

The mixing of quantum fields plays an important role in the phenomenology
of high-energy physics \cite{1,18,15}.
Mixing of both $K^0\bar K^0$ and $B^0\bar B^0$
bosons provides the evidence of $CP$ violation in weak interaction 
\cite{14}
and $\eta\eta \prime $ boson mixing in the $SU(3)$ flavor group gives a
unique opportunity to investigate the nontrivial QCD vacuum and fill
the gap between QCD and the constituent quark model. In the fermion
sector, neutrino mixing and oscillations provide striking evidences of 
neutrino masses and are the most likely solution of the
famous solar neutrino puzzle \cite{16,17,26}. In addition, the standard 
model
incorporates the mixing of fermion fields through the
Kobayashi-Maskawa (CKM) mixing of three quark flavors, a generalization
of the original Cabibbo mixing matrix between $d$ and $s$ quarks
\cite{11,22,24,29}.

From the theoretical point of view the field theory of mixing 
is important since this is one of a few examples where the quantum field
theory can be solved exactly.
Moreover, the field-theoretical treatment of mixing touches the fundamental
questions of the
quantization of interacting fields, which are not yet fully understood. 
It also allows to improve the accuracy of ordinary
perturbation theory, e.g. the mixed-Hamiltonian dynamics 
can be used for partial re-summation of ordinary perturbation
series in weak interactions.

Recently, the importance of mixing transformations has prompted
a fundamental examination from a field-theoretical 
perspective.
It was found that the flavor mixing in quantum field theory
introduces very non-trivial relationships between the interacting 
and non-interacting (free) fields,
which lead to {\em  unitary inequivalence} between the Fock space 
of the interacting fields and that of the free fields \cite{3,4,19,20}.
This is different
from the conventional 
perturbation theory where one would expect the vacuum of 
the interacting fields to be essentially similar to one of the free fields 
(up to a phase factor $e^{iS_0}$ \cite{21,10}). 
The investigation of two-field unitary mixing in the fermion sector
by Blasone and Vitiello \cite{4,5,8,9}
demonstrated a rich structure of the interacting-field vacuum as SU(2)
coherent state and altered the oscillation formula to include the
antiparticle degrees of freedom.
Subsequent analysis of the boson case revealed a similar
but much richer structure of the vacuum of the interacting fields \cite{3,23,7}. 
Especially, the pole structure in the inner
product between the vacuum of the free theory and the vacuum of the 
interacting theory was found and
related to the convergence radius of the perturbation series \cite{23}.
Attempts to look at the mixing of more than two flavors have also been 
carried out and a general framework for such a theory has been 
suggested by Ji and Mishchenko \cite{JM}. 
Also, mathematically rigorous study of this case has been offered by Hannabuss
and Latimer \cite{HAN}.

Surprisingly, although a wide research effort has been undertaken 
in this direction,
an important aspect of the formulation is still left uncultivated.
Namely, the oscillation formulas in most cases are 
obtained theoretically in terms of {\em time oscillations}, 
while the experimental measurements are only carried out in 
terms of {\em space oscillations}.
Despite seeming triviality of the question, it is indeed not an easy one and
certain controversy always existed in the literature about the appropriate way
of handling this task. Naively speaking, a quantity of dimension of speed must
be involved in the conversion, however there are at least two such 
quantities in the flavor mixing problem \cite{Lipkin}.
It was argued that the correct way
to do such conversion was to equal the energies of the mixed particles thus 
allowing the uncertainty in 3-momenta to form a wave-packet.
At the higher level of rigor, no doubt, 
the most consistent argument in quantum mechanics comes from the method
of wave-packets, where the flavor 
particles are treated as waves with not
just uncertainty in the 3-momentum, but also in the energy \cite{30,Giunti}.
Within this approach the correct {\em oscillation length} 
is recovered and also 
the concept of {\em correlation length} is introduced. 
As one becomes aware of nontrivial role of antiparticles in 
field-theoretical flavor mixing \cite{JM,4}, a natural question arises if and how
such effects would manifest themselves in space-dimensions.
Therefore, in this work we would like 
to build a field-theoretical development 
analogous to quantum mechanical wave-packet method
and apply it to study time-to-space conversion
for the field-theoretical result in flavor mixing.

We must note that recently the wave-packet approach has been extended to 
field theory 
by means of using the field-theoretical flavor
propagation functions \cite{Beuthe}. We should stress, however, that such treatment 
is essentially different from ours in that the flavor propagators were 
defined via the expectation value on the vacuum of the free-fields 
and as such have the problem of total probability non-conservation 
pointed out by Blasone and Vitiello \cite{6,5}.
It thus lacks the capacity to include nonperturbative effects due
to nontrivial flavor condensation in vacuum found in
\cite{JM,3,23,7,6,4,5}.

In this paper, we analyze the problem of time-to-space conversion from the 
same "wave-packets" point of view, but entirely within the canonical framework of
creation/annihilation operators. This allows us to build
the theory as an extension of the general formalism \cite{JM} and, thus, 
to consider the nontrivial vacuum effects. 
In particular, the high-frequency 
contribution on top of usual Ponte-Corvo 
oscillations is retained and its effect in space-dimensions is studied. 

In the next section (Section II) we present the canonical framework for
the flavor oscillations in space-time and generalize the quantum
mechanical wave-packet method to the quantum field theory. In Section III, as
a demonstration of this formalism, we discuss the 
space-oscillations in the system of two (pseudo)scalar mesons such as $\eta$ 
and $\eta^\prime$ and present the relevant numerical results.
Conclusion follows in Section IV. 

\section{Oscillations of flavor in space-time}

We begin our study with a simple example. 
Let us consider a particle created initially in quantum
state $\left| i\right\rangle $ and propagated in space and time.
The number of particles of 
sort $\rho $ to be detected at the space-time position $x=(t,{\bf x})$ is
given by 
\begin{equation}
\label{eqn001a}N_\rho(x)=\left\langle i|a_\rho ^{\dagger }(x)
a_\rho(x)|i\right\rangle ,
\end{equation}
where $a^\dag_\rho (x)(a_\rho (x)) $ is creation (annihilation) operator for
particles of sort $\rho$ at 
space-time position $x$. This can be defined via 
creation (annihilation) operators for given momentum ${\bf k}$
\begin{equation}
\label{eqn002a}a_\rho ^{\dagger }(t,{\bf x})=\sum\limits_{{\bf k}}\frac 1{
\sqrt{2\omega _{\rho {\bf k}}}}e^{-i{\bf kx}}
a_{\rho ,{\bf k}}^{\dagger}(t). 
\end{equation}
Substituting the definition given by 
Eq.(\ref{eqn002a}) into Eq.(\ref{eqn001a}), one obtains 
\begin{equation}
\label{eqn003a}
N_\rho(x)=\sum\limits_{{\bf k,k}^{\prime }}
\frac{e^{i({\bf k}^{\prime}-{\bf k}){\bf x}}}
{2\sqrt{\omega _{\rho {\bf k}}
\omega_{\rho{\bf k}^{\prime }}}}
\left\langle i|a_{\rho ,{\bf k}}^{\dagger }(t)
a_{\rho ,{\bf k}^{\prime }}(t)|i\right\rangle . 
\end{equation}
We thus find that the number of particles expected at the space-time 
position $x$ can be found using Eq.(\ref{eqn003a}) once 
$\left\langle i|a_{\rho ,{\bf k}}^{\dagger }(t)
a_{\rho ,{\bf k}^{\prime}}(t)|i\right\rangle $ 
is known for all ${\bf k}$ and ${\bf k}^{\prime }$.

In the above example, it is instructive to recognize a more
general problem. First of all, note that Eq.(\ref{eqn003a}) is 
analogous to conventional probability density 
$|\Psi(x)|^2$; in the case of free fields it, indeed, yields 
the square of the Feynman propagation amplitude
\begin{equation}
\label{eqn004c}
|\Delta _F(z)|^2=\left| 
\sum\limits_{{\bf k}}
\frac{e^{i({\bf k z}-\omega _{{\bf k}}z^0)}}{2\omega _{{\bf k}}}\right|^2
=|\int \frac{d^d k}{(2\pi)^d} \frac{i e^{-i k z}}{k^2-m^2+i\epsilon}|^2. 
\end{equation}
Unlike the Feynman propagation amplitude, however, Eq.(\ref{eqn003a})
generalizes naturally to the case when the total
number of particles of a given sort
is not conserved, as is,
essentially, in any flavor oscillation theory.

Now, following closely the structure of the states in field theory of flavor
mixing \cite{JM,7,4}, let us introduce the initial 
state $\left|i\right\rangle $
for a system with $N$ sorts (flavors) of particles by 
\begin{equation}
\label{eqn004d}
\left| i\right\rangle =
\sum\limits_{\rho=1;{\bf k}}^N
(g_{\rho ,{\bf k}}a_{\rho ,{\bf k}}^{\dagger }+h_{\rho ,{\bf k}}
b_{\rho ,-{\bf k}}^{\dagger })\left| \Omega\right\rangle, 
\end{equation}
where $a_{\rho ,{\bf k}}^{\dagger }$, $b_{\rho ,-{\bf k}}^{\dagger }$
are the particle and antiparticle creation operators for sort (flavor) $\rho$
and $\left| \Omega\right\rangle$ is the flavor vacuum state annihilated by 
all $a_{\rho ,{\bf k}}$, $b_{\rho ,{\bf k}}$. 
Eq.(\ref{eqn004d}), essentially, represents a single particle 
initially created in a state such that the probability 
to observe it as sort $\rho$-particle
(antiparticle) is simply  $|g_{\rho ,{\bf k}}|^2$ ($|h_{\rho ,-{\bf k}}|^2$).
Then, $2N$ functions $g_{\rho,{\bf k}}, h_{\rho,{\bf k}}$ 
are the form-factors for the initial state $\left| i\right\rangle$
\footnote{In Eq.(\ref{eqn004d}) we assumed that 
particles and antiparticles are distinguishable. Although this is feasible in 
the case of, e.g., neutrinos, for many cases of meson mixing the 
field operators are 
self-adjoint and thus particles may not be distinguished from antiparticles.
However, Eq.(\ref{eqn004d}) can still be used by redefining
$b^\dagger\equiv a^\dagger$ and $g_{\rho,{\bf k}}\equiv h_{\rho,-{\bf k}}$.
Given this remark, we will continue with general formulation keeping in mind
that the meson-mixing can be obtained with straightforward adjustments from
our final results.}.
For notation convenience we shall adopt following convention. We will let
$\alpha $ be both positive and negative ($\alpha
=-N,\ldots ,N$, excluding $\alpha=0$) 
with negative $\alpha $ enumerating antiparticles and
positive $\alpha $ enumerating particles, respectively. In this notation 
\begin{equation}
\label{eqn005a}
\left| i\right\rangle =
\sum\limits_{\alpha=-N;{\bf k}}^N
f_{\alpha ,{\bf k}}a_{\alpha ,{\bf k}}^{\dagger }\left| \Omega\right\rangle, 
\end{equation}
where for convenience we have introduced 
$a_{\alpha,{\bf k}}^{\dag}:=a^\dag_{\alpha,{\bf k}}$
{\em for} $\alpha>0$ and $a_{\alpha,{\bf k}}^\dag:=b^\dag_{-\alpha,-{\bf k}}$
{\em for} $\alpha<0$. 
Analogously, $f_{\alpha,{\bf k}}:=g_{\alpha,{\bf k}}$ 
{\em for} $\alpha>0$ and $f_{\alpha,{\bf k}}:=h_{-\alpha,{\bf k}}$ 
{\em for} $\alpha<0$. Also, from now on,
in summations over $\alpha$ we imply $\alpha\neq 0$.

From Eq.(\ref{eqn005a}), we may further introduce 
the creation operator for form-factor 
${ F}=\{f_{\alpha,{\bf k}},\alpha=-N\dots -1, 1\dots N\}$ as 
\begin{equation}
a^{\dagger }({ F})=\sum\limits_{\alpha =-N;{\bf k}}^N
f_{\alpha ,{\bf k}}a_{\alpha ,{\bf k}}^{\dagger }, 
\end{equation}
so that concisely $\left|i\right\rangle =a^{\dagger }({F})\left| \Omega\right\rangle$. 
It is straightforward to obtain the non-equal time commutation/anticommutation relationships
for $a({F})$ and $a^{\dagger }(G)$
\begin{equation}
\begin{array}{c}
\lbrack a({F}),a^{\dagger }({G})]_
{\pm , \text{equal time}}
=\sum\limits_{\alpha =-N;{\bf k}}^Nf_{\alpha ,{\bf k}}^{*}
g_{\alpha ,{\bf k}}=(f,g)_I, \\ 
\lbrack a_t({F}),a^{\dagger }({G})]_{\pm}=
\sum\limits_{\alpha,\beta =-N;{\bf k}}^N
f_{\alpha ,{\bf k}}^{*}{\cal F}_{\alpha \beta ;{\bf k}}(t)
g_{\beta ,{\bf k}}=(f,g)_{\cal F}, 
\\ \lbrack a_t({F}),a({G})]_{\pm}=
\sum\limits_{\alpha ,\beta =-N;{\bf k}}^Nf_{\alpha ,{\bf k}}^{*}
{\cal G}_{\alpha \beta ;{\bf k}}(t)g^*_{\beta ,{\bf k}}=(f,g^*)_{\cal G},
\end{array}
\end{equation}
where $\pm$ in $[$ $]_\pm$ corresponds to commutation/anticommutation and
the inner product is defined by
$(f,g)_{\cal A}=\sum\limits_{\alpha ,\beta =-N;{\bf k}}^N
f_{\alpha ,{\bf k}}^{*}{\cal A}_{\alpha \beta ;{\bf k}}(t)g_{\beta ,{\bf k}}$.
${\cal F}_{\alpha \beta ;{\bf k}}(t)$, ${\cal G}_{\alpha \beta ;{\bf k}}(t)$
are the non-equal time commutators/anticommutators for momentum ${\bf k}$
as introduced in \cite{JM},
\begin{equation}
\begin{array}{c}
[a_{\alpha,{\bf k}}(t),a^\dag_{\beta,{\bf k'}}]_\pm =
{\cal F}_{\alpha \beta; {\bf k}}(t) \delta({\bf k - k'}) \cr
[a_{\alpha,{\bf k}}(t),a_{\beta,{\bf k'}}]_\pm =
{\cal G}_{\alpha \beta; {\bf k}}(t)\delta({\bf k - k'}).
\end{array}
\end{equation}
In the notations used in our general field theory of 
flavor mixing \cite{JM}, for given momentum {\bf k},
\begin{equation}
\begin{array}{c}
{\cal G}_{\alpha \beta }(t)=\left\{ 
\begin{array}{c}
G_{-\alpha ,\beta }(t),\alpha <0,\beta >0; \\ 
\eta G_{-\beta ,\alpha }(-t),\alpha >0,\beta <0; \\ 
0,\text{ otherwise.} 
\end{array}
\right. =\left( 
\begin{array}{cc}
0 & \eta G^T(-t) \\ 
G(t) & 0 
\end{array}
\right) , \\ 
{\cal F}_{\alpha \beta }(t)=\left\{ 
\begin{array}{c}
F_{\alpha ,\beta }(t),\alpha >0,\beta >0; \\ 
F_{-\beta ,-\alpha }(t),\alpha <0,\beta <0; \\ 
0,\text{ otherwise.} 
\end{array}
\right| =\left( 
\begin{array}{cc}
F(t) & 0 \\ 
0 & F^T(t) 
\end{array}
\right) , 
\end{array}
\end{equation}
where $\eta=(-1)^{2S}$ with $S$ being the spin of the mixed fields
($\eta$ is +1 for bosons and -1 for fermions), and 
$F_{\alpha\beta}$ and 
$G_{\alpha\beta}$ in terms of original flavor creation/annihilation
operators are defined by
\begin{equation}\label{eqn001x}
\begin{array}{rl}
F_{\alpha \beta ; 
{\bf k}}(t)
&=[a_{\alpha ,{\bf k}}(t),a_{\beta ,{\bf k}}^{\dagger }]_{\pm} \\
&=[b_{\beta {\bf ,-k}}(t),b_{\alpha {\bf ,-k}}^{\dagger }]_{\pm }, \\ 
G_{\alpha \beta ;{\bf k}}(t)
&=[b_{\alpha ,-{\bf k}}(t),a_{\beta ,{\bf k}}]_{\pm }. 
\end{array}
\end{equation}

The primary interest of this paper is the process in which a "flavor" particle
is created in some initial state $|i\rangle$ with form-factor
given by $G=\{g_{\alpha,{\bf k}}\}$ and is detected at time $t$ as state
$|f\rangle$ with form-factor $F=\{f_{\alpha,{\bf k}}\}$.
In that case the object of interest is  the number operator 
$N_F(t)=a_t^{\dagger}(F)a_t(F)$
\begin{equation}
\label{eqn012a}
\left\langle i|N_F(t)|i\right\rangle =
\left\langle\Omega|a(G)a_t^{\dagger }(F)a_t(F)a^{\dagger }(G)|\Omega\right\rangle. 
\end{equation}
Using the machinery we just have introduced and after simple algebra, we find 
\begin{equation}
\label{eqn013a}
\begin{array}{rl}
\left\langle i|N_F(t)|i\right\rangle & 
=|[a_t(F),a^{\dagger }(G)]_{\pm}|^2
+\eta |[a(G),a_t(F)]_{\pm }|^2
+\left\langle a_t^{\dagger}(F)a_t(F)\right\rangle , \\ 
& 
=|(f,g)_{\cal F}|^2+
\eta|(f,g^*)_{\cal G}|^2+(f,f)_{\cal Z}, 
\end{array}
\end{equation}
where the last term is related to flavor condensation in vacuum
\begin{equation}
\begin{array}{c}
(f,f)_{\cal Z}=\left\langle a_t^{\dagger }(F)a_t(F)\right\rangle =
\sum\limits_{\alpha =-N;{\bf k}}^N
f_{\alpha ,{\bf k}}{\cal Z}_{\alpha ,{\bf k}}(t)f_{\alpha ,{\bf k}}, \cr
{\cal Z}_{\alpha ,{\bf k}}(t)=
\left\langle \Omega|a_{\alpha ,{\bf k}}^{\dagger }(t)
a_{\alpha ,{\bf k}}(t)|\Omega\right\rangle .
\end{array}
\end{equation}
Generally, this term is not zero for $t\neq 0$ as was found in the works
mentioned above. Indeed, for many choices of $F$ it is infinite.
On the other hand, one may notice that this contribution is independent
from the initial state $|i\rangle$
and thus may be interpreted as
the background due to the vacuum
condensation picked up by the detector itself. In this case
one shall renormalize it by defining
$$
\left\langle i|N_F(t)|i\right\rangle _r=
\left\langle i|N_F(t)|i\right\rangle -
\left\langle \Omega| N_F(t)| \Omega \right\rangle.
$$ 
In what follows, we will understand $\left\langle i|N_F(t)|i\right\rangle$ by such
a renormalized expectation value. In an extended notation, explained above, 
it is given by
\begin{equation}\label{eqn014a}
\begin{array}{c}
\left\langle i|N_F(t)|i\right\rangle_r =
|\sum\limits_{\alpha ,\beta =-N;{\bf k}}^N
f_{\alpha ,{\bf k}}^{*}
{\cal F}_{\alpha \beta ;{\bf k}}(t)g_{\beta ,{\bf k}}|^2+
\eta|\sum\limits_{\alpha ,\beta=-N;{\bf k}}^N
f_{\alpha ,{\bf k}}^{*}
{\cal G}_{\alpha \beta ;{\bf k}}(t)g^*_{\beta ,{\bf k}}|^2.
\end{array}
\end{equation}
In the previous works special attention was paid to the oscillations of flavor charge
which can be immediately related to the observables in the theory. 
Oscillations of flavor charge can be 
obtained from Eq.(\ref{eqn013a}). For the case
of detection of a single particle with flavor $\beta$ and form-factor $f(\bf k)$
\begin{equation}\label{eqnQ}
Q_\beta (t) = N_{[\beta]}(t) - N_{{[-\beta]}}(t), 
\end{equation}
where 
$[\beta]=\{f_{\alpha,{\bf k}}=f({\bf k})\delta_{\alpha, \beta }\}$ 
and ${[-\beta]}=\{f_{\alpha,{\bf k}}=f({-\bf k})\delta_{-\alpha, \beta }\}$.

It is straightforward to derive from Eqs.(\ref{eqn014a}) and (\ref{eqnQ})
the oscillations in space, in which case the detector shall be 
characterized by the form-factor;
$[\beta]=\{f_{\alpha, {\bf k}} = 
\frac{e^{-i{\bf k x}}}{\sqrt{2\omega_{\beta {\bf k}}}}
\delta_{\alpha ,\beta }\}$ for a single particle of sort $\beta$
or
$[-\beta]=\{f_{\alpha, {\bf k}} = 
\frac{e^{i{\bf k x}}}{\sqrt{2\omega_{\beta {\bf k}}}}
\delta_{-\alpha ,\beta }\}$ for a single anti-particle of the same sort:
\begin{equation}
\label{eqn015a}
\begin{array}{rl}
\left\langle i|N_\beta(t,{\bf x})|i\right\rangle _r
=&|\sum\limits_{\alpha;{\bf k}}
\frac{e^{i{\bf kx}}}{({2\omega_{\beta {\bf k}}})^{1/2}}{{\cal F}_{\beta \alpha ;{\bf k}}(t)g_{\alpha ,{\bf k}}}|^2+
\eta |\sum\limits_{\alpha;{\bf k}}
\frac{e^{i{\bf kx}}}{({2\omega_{\beta {\bf k}}})^{1/2}}{{\cal G}_{\beta \alpha ;{\bf k}}(t)g^*_{\alpha ,{\bf k}}}|^2, \\ 
\left\langle i|N_{-\beta}(t,{\bf x})|i\right\rangle _r
=&|\sum\limits_{\alpha;{\bf k}}
\frac{e^{-i{\bf kx}}}{({2\omega_{\beta {\bf k}}})^{1/2}}{{\cal F}_{-\beta \alpha ;{\bf k}}(t)g_{\alpha ,{\bf k}}}|^2+
\eta |\sum\limits_{\alpha;{\bf k}}
\frac{e^{-i{\bf kx}}}{({2\omega_{\beta {\bf k}}})^{1/2}}{{\cal G}_{-\beta \alpha ;{\bf k}}(t)g^*_{\alpha ,{\bf k}}}|^2, \\ 
\left\langle i|Q_\beta(t,{\bf x})|i\right\rangle =&
\left\langle i|N_{\beta}(t,{\bf x})|i\right\rangle-
\left\langle i|N_{-{\beta}}(t,{\bf x})|i\right\rangle . 
\end{array}
\end{equation}
As in the quantum-mechanical wave-packet method, our final result 
depends on the kind of initial state assumed for the flavor particle. 
For example, one may consider the initial state as a state with
definite momentum ${\bf k}$ and definite flavor $\beta$, 
\begin{equation}
\label{eqn018a}
\left\langle \beta {\bf k}|N_\alpha(t,{\bf x})|\beta {\bf k}\right\rangle _r=
(2\omega_{\alpha {\bf k}}\omega _{\beta {\bf k}})^{-1}{|{\cal F}_{\alpha \beta ;{\bf k}}(t)|^2}+
\eta
(2\omega_{\alpha {\bf k}}\omega _{\beta {\bf k}})^{-1}{|{\cal G}_{\alpha\beta ;{\bf k}}(t)|^2}. 
\end{equation}
This, however, has no dependence on ${\bf x}$ and thus one can not observe any space
oscillation. One also might consider a particle of sort $\beta $
created at position ${\bf x}^{\prime }$ and observed at position ${\bf x}$
at time $t$ as a particle of sort $\alpha$. For that case one finds
\begin{equation}
\label{eqn017a}
\begin{array}{c}
\left\langle \beta {\bf x}^{\prime }|N_\alpha(t,{\bf x})
|\beta {\bf x}^{\prime }\right\rangle_r=
|\sum\limits_{{\bf k}}
\frac{e^{i({\bf x}-{\bf x}^{\prime }){\bf k}}}
{2\sqrt{\omega_{\alpha {\bf k}}\omega_{\beta {\bf k}}}}{{\cal F}_{\alpha \beta ;{\bf k}}(t)}|^2+
\eta |\sum\limits_{{\bf k}}
\frac{e^{i({\bf x}-{\bf x}^{\prime }){\bf k}}}
{2\sqrt{\omega_{\alpha {\bf k}}\omega_{\beta {\bf k}}}}{{\cal G}_{\alpha \beta ;{\bf k}}(t)}|^2 \\ 
=|^{*}{\cal F}_{\alpha \beta }({\bf x}^{\prime }-{\bf x},t)|^2+
\eta|^{*}{\cal G}_{\alpha \beta }({\bf x}^{\prime }-{\bf x},t)|^2, 
\end{array}
\end{equation}
where $^{*}{\cal F}({\bf z})$ is the Fourier transform of 
$(2\sqrt{\omega_{\alpha {\bf k}}\omega _{\beta {\bf k}}})^{-1}{{\cal F}({\bf k})}$ 
and $^*{\cal G}$ is defined similarly. 

In practice, however, we are interested in the case when 
a flavor particle was produced originally in a small (but finite) region of space
with (nonzero) momentum ${\bf k}$.
This can be represented by a 
well-peaked initial state $|i \rangle$ with form-factor
$g({\bf k})$
such that a single particle of sort $\beta$ appears as a
wave-packet of momentum ${\bf k}_0$ with small dispersion $\sigma $. 
Taking now the explicit form of ${\cal F}$ and ${\cal G}$ 
from \cite{JM} and
leaving the detector point-like, we obtain
\begin{equation}
\label{eqn019b}
\begin{array}{c}
\left\langle \beta g|N_\alpha(t,{\bf x})|\beta g\right\rangle_r =
\left| \sum\limits_{\gamma }(a_{\alpha\beta ;\gamma }
e^{-iw_\gamma t}e^{-\sigma ^2({\bf v}_\gamma t-{\bf x})^2/2}+
b_{\alpha \beta ;\gamma }e^{iw_\gamma t}
e^{-\sigma ^2(-{\bf v}_\gamma t-{\bf x})^2/2})\right|^2 + \\ 
\eta \left| \sum\limits_{\gamma }(c_{\alpha \beta;\gamma }
e^{-iw_\gamma t}e^{-\sigma ^2( {\bf v}_\gamma t-{\bf x})^2/2}+
d_{\alpha \beta ;\gamma }e^{iw_\gamma t}
e^{-\sigma^2(-{\bf v}_\gamma t-{\bf x})^2/2})\right|^2 ; \\ 
{\bf v}_\gamma =\frac {d w_\gamma ({\bf k})}{d{\bf k}}|_{{\bf k}_0} .
\end{array}
\end{equation}
In the above derivation we used following identity which 
can be proved
using stationary phase approximation
for
function $g({\bf k})$
sharply peaked around ${\bf k=k}_0$ and slow-varying functions
$f({\bf k})$ and $S({\bf k})$, 
\begin{equation}
\begin{array}{c}
\int d {\bf k}
g({\bf k})f({\bf k})e^{iS({\bf k})}e^{i{\bf k x}}\approx 
(2\pi\sigma^2)^{3/2}
g({\bf k}_0)f({\bf k}_0)e^{i(S({\bf k}_0)+{\bf k}_0{\bf x})}
\exp(-\sigma ^2({\bf x}+{\bf \vec\nabla }S({\bf k}_0))^2/2), \\ 
\sigma^2=-\frac{g({\bf k}_0)}{g^{\prime \prime }({\bf k}_0)}.
\end{array}
\end{equation}
The explicit form of ${\cal F}$ and ${\cal G}$ is taken as
(see \cite{JM} for specific values of 
$a_{\alpha,\beta;\gamma}({\bf k}),b_{\alpha,\beta;\gamma}({\bf k}),
c_{\alpha,\beta;\gamma}({\bf k}),d_{\alpha,\beta;\gamma}({\bf k})$)
\begin{equation}
\label{eqn018b}
\begin{array}{c}
({\sqrt{2\omega _{\alpha {\bf k}}}})^{-1}{{\cal F}_{\alpha \beta ,{\bf k}}(t)}
=\sum\limits_{\gamma=1}^N
(a_{\alpha \beta ;\gamma }({\bf k})e^{-iw_{\gamma {\bf k}}t}+
b_{\alpha \beta ;\gamma }({\bf k})e^{iw_{\gamma {\bf k}}t}), \\ 
({\sqrt{2\omega _{\alpha {\bf k}}}})^{-1}{{\cal G}_{\alpha\beta ,{\bf k}}(t)}
=\sum\limits_{\gamma=1}^N
(c_{\alpha \beta ;\gamma }({\bf k})e^{-iw_{\gamma {\bf k}}t}+
d_{\alpha \beta;\gamma }({\bf k})e^{iw_{\gamma {\bf k}}t}), 
\end{array}
\end{equation}
and all amplitudes in Eq.(\ref{eqn019b})
are taken at ${\bf k=k}_0$. 
Eq.(\ref{eqn019b}) 
physically represents the expectation value for the
number of $\alpha $-sort particles observed at position
${\bf x}$ at time $t$ when a
single particle of sort $\beta $ and form-factor $g({\bf k})$ had been emitted. 
What we observe, hence, is a single wave-packet propagating
through the space: i.e. one can see that 
$\left\langle \beta g|N_t(\alpha {\bf x})|\beta g\right\rangle $ 
reaches maximum only when the
"center" of the wave-packet passes over the observation point ${\bf x}$
\begin{equation}
\langle{\bf v}\rangle t\approx {\bf x}\text{ or }
\langle-{\bf v}\rangle t\approx {\bf x}. 
\end{equation}
To explicitly observe space oscillations, one should take an average 
over time in Eq.(\ref{eqn015a})
which would correspond to a 
observation continuous in time,
\begin{equation}
\label{eqn019a}
\begin{array}{rl}
W_g(f)&\sim \lim \limits_{T\rightarrow \infty } 
\int\limits_Tdt\left\langle g|N(t,F)|g\right\rangle_r  \\ 
&\sim\lim\limits_{T\rightarrow \infty }
\int\limits_Tdt\left( |\sum\limits_{\beta;{\bf k}}
\frac{e^{i{\bf kx}}}{\sqrt{2\omega _{\alpha {\bf k}}}}
{\cal F}_{\alpha \beta ;{\bf k}}(t)g_{\beta ,{\bf k}}|^2+\eta
|\sum\limits_{\beta;{\bf k}}
\frac{e^{i{\bf kx}}}{\sqrt{2\omega _{\alpha {\bf k}}}}
{\cal G}_{\alpha \beta ;{\bf k}}(t)g^*_{\beta ,{\bf k}}|^2\right) . 
\end{array}
\end{equation}
Using Eq.(\ref{eqn018b}), we may rewrite this as
$$
\begin{array}{c}
\sum\limits_{\gamma ,\gamma ^{\prime }}\int \int d 
{\bf k}d{\bf k}^{\prime }\delta (w_{\gamma {\bf k}}-
w_{\gamma ^{\prime }{\bf k}^{\prime }})
[(a_{\alpha \beta ;\gamma }({\bf k})
a_{\alpha \beta ;\gamma^{\prime }}^{*}({\bf k}^{\prime })+
b_{\alpha \beta ;\gamma }({\bf k})
b_{\alpha \beta ;\gamma ^{\prime }}^{*}({\bf k}^{\prime }))+ \\ 
\eta (c_{\alpha \beta ;\gamma }({\bf k})c_{\alpha \beta ;\gamma ^{\prime }}^{*}
({\bf k}^{\prime })+
d_{\alpha \beta ;\gamma }({\bf k})
d_{\alpha \beta ;\gamma^{\prime }}^{*}({\bf k}^{\prime }))]
g_{\beta,{\bf k}}g^{*}_{\beta,{\bf k}^{\prime }}
e^{i({\bf k}-{\bf k}^{\prime }){\bf x}} .
\end{array}
$$
When the mass difference $m_\gamma^2-m_{\gamma'}^2$ is small, the functional 
\begin{equation}
\begin{array}{c}
\delta (w_{\gamma 
{\bf k}}-w_{\gamma ^{\prime }{\bf k}^{\prime }})[(a_{\alpha \beta ;\gamma }(
{\bf k})a_{\alpha \beta ;\gamma ^{\prime }}^{*}({\bf k}^{\prime })+b_{\alpha
\beta ;\gamma }({\bf k})b_{\alpha \beta ;\gamma ^{\prime }}^{*}({\bf k}^{\prime }))\pm \\ 
(c_{\alpha \beta ;\gamma }({\bf k})c_{\alpha \beta;\gamma ^{\prime }}^{*}({\bf k}^{\prime })+
d_{\alpha \beta ;\gamma }({\bf k})d_{\alpha \beta ;\gamma ^{\prime }}^{*}({\bf k}^{\prime }))]
g_{\beta,{\bf k}}g^{*}_{\beta,{\bf k}^{\prime }}e^{i({\bf k}-{\bf k}^{\prime }){\bf x}} 
\end{array}
\end{equation}
has its maximum at 
${\bf k}\approx{\bf k'}\approx {\bf k}_0$.
Then for the initial single flavor particle 
$\beta$ with sharply-peaked $g_{\beta,{\bf k}}$ we 
can use the stationary phase approximation again
to find
\begin{equation}
\begin{array}{rll}
W({\bf x})&\sim
\sum\limits_{\gamma ,\gamma ^{\prime }}
&\left[(a_{\alpha \beta ;\gamma }({\bf k})a^*_{\alpha \beta ;\gamma' }({\bf k})+
b_{\alpha \beta ;\gamma}({\bf k})b^*_{\alpha \beta ;\gamma'}({\bf k}))\right.\cr
&&+ 
\left.\eta (c_{\alpha \beta ;\gamma }({\bf k})c^*_{\alpha \beta ;\gamma' }({\bf k})+
d_{\alpha \beta;\gamma }({\bf k})d^*_{\alpha \beta;\gamma' }({\bf k}))\right]
e^{-(\frac{\Delta k_{\gamma\gamma'}}{2 k_0})^2\sigma^2{\bf x}^2+
i\Delta {\bf k}_{\gamma\gamma'}{\bf x}},
\end{array} 
\end{equation}
where ${\bf k}\approx {\bf k}_0$, 
$\Delta {\bf k}_{\gamma \gamma'}=
\frac{\Delta m^2_{\gamma \gamma'}}{2|{\bf k_0}|}\hat{\bf k}_0=
\frac{m^2_{\gamma}-m^2_{\gamma'}}{2|{\bf k}_0|}\hat{\bf k}_0$. 

For the mixing of two flavors 
we recover the oscillation length as 
\begin{equation}
\label{eqn022a}L\approx \frac{2k}{\Delta m_{12}^2}. 
\end{equation}
We should point out here that the field-theoretical corrections are found
only to change the amplitude of the oscillations while no major
distortion in the structure of oscillations is found. This is quite
different from the case
of the flavor oscillations in time where the additional high-frequency
term was prominent. Indeed, we found that only one mode of
space oscillations, with the wave-length related
to $\Delta m_{\gamma \gamma ^{\prime }}^2$ by Eq.(\ref{eqn022a}), 
survives
the average over time. While this result may be in part
attributed to the approximation we used, one may also notice that
in Eq.(\ref{eqn019a})
$$
\int dt e^{i (\omega_{\gamma{\bf k}}\pm\omega_{\gamma'{\bf k'}}) t}
$$
is not equal to zero only when 
$\omega_{\gamma{\bf k}}\pm\omega_{\gamma'{\bf k'}}=0$ and for that the
frequencies must come in with the opposite signs. Thus, no high-frequency
terms may survive the integration over time.
However, field-theoretical effect is still noticeable in the shape 
of the oscillations as we will show below.

\section{Space-oscillations in mixing of two mesons}

In this section we shall investigate in greater details the effect of the
field-theoretical corrections on the shape of the oscillations of flavor charge in space. 
For that, we will apply the formalism developed above to the mixing of two
spin-zero particles created initially in a state
with given flavor and a gaussian form-factor.

In the case of two spin-zero fields the mixing can be described completely 
with a single
mixing angle which relates the flavor fields ($\varphi $) to the
free fields ($\phi $) by
\begin{equation}
\left( 
\begin{array}{c}
\varphi _1 \\ 
\varphi _2
\end{array}
\right) =\left( 
\begin{array}{cc}
\cos (\theta ) & -\sin (\theta ) \\ 
\sin (\theta ) & \cos (\theta )
\end{array}
\right) \left( 
\begin{array}{c}
\phi _1 \\ 
\phi _2
\end{array}
\right) .
\end{equation}
The time evolution in this system has been studied 
in quantum field theoretical framework in \cite{23,7}. 
It was found that, with the account of nontrivial vacuum effect, 
the formula for the flavor charge oscillations
should change and acquire the additional high-frequency term, i.e.
\begin{equation}
\begin{array}{c}
Q_1=1+\sin ^2(\theta )(\gamma _{-}^2\sin ^2(\Omega)-\gamma _{+}^2\sin^2(\omega)), \\ 
Q_2=\sin ^2(\theta )(\gamma _{+}^2\sin ^2(\omega)-\gamma _{-}^2\sin^2(\Omega))
\end{array}
\end{equation}
with $\omega =(\omega _{1{\bf k}}-\omega _{2{\bf k}})/2$ and 
$\Omega =(\omega _{1{\bf k}}+\omega _{2{\bf k}})/2$, 
$\gamma _{\pm }=(\sqrt{\omega _{1{\bf k}}/\omega _{2{\bf k}}}\pm 
\sqrt{\omega _{2{\bf k}}/\omega _{1{\bf k}}})/2$. It
was also found that the time evolution of the flavor fields can be described by
non-equal time commutation relations \cite{JM,23}
(see Eq.(\ref{eqn001x}))
\begin{equation}
\begin{array}{l}
F_{11;{\bf k}}(t)=
\cos ^2(\theta )e^{-i\omega _{1{\bf k}}t}+
\gamma _{+}^2\sin^2(\theta )e^{-i\omega _{2{\bf k}}t}-
\gamma _{-}^2\sin ^2(\theta )e^{i\omega_{2{\bf k}}t}, \\ 
F_{12;{\bf k}}(t)=
F_{21;{\bf k}}(t)=\gamma _{+}\sin (\theta )\cos (\theta)
(e^{-i\omega _{2{\bf k}}t}-e^{-i\omega _{1{\bf k}}t}), \\ 
F_{22;{\bf k}}(t)=
\cos ^2(\theta )e^{-i\omega _{2{\bf k}}t}+
\gamma _{+}^2\sin ^2(\theta)e^{-i\omega _{1{\bf k}}t}-
\gamma _{-}^2\sin ^2(\theta )e^{i\omega _{1{\bf k}}t}
\end{array}
\end{equation}
and
\begin{equation}
\begin{array}{l}
G_{11;{\bf k}}(t)=
\gamma _{+}\gamma _{-}\sin ^2(\theta )(e^{-i\omega _{2{\bf k}}t}-
e^{i\omega _{2{\bf k}}t}), \\ 
G_{12;{\bf k}}(t)=
-G_{21;{\bf k}}(-t)=\gamma _{-}\sin (\theta )\cos (\theta)
(e^{-i\omega _{1{\bf k}}t}-e^{i\omega _{2{\bf k}}t}), \\ 
G_{22;{\bf k}}(t)=
\gamma _{+}\gamma _{-}\sin ^2(\theta )(e^{-i\omega _{1{\bf k}}t}-e^{i\omega_{1{\bf k}}t}).
\end{array}
\end{equation}

We will consider the time evolution of a single particle born in 
flavor $\alpha=1$ with a sharp
gaussian form-factor centered at the average momentum 
{\bf k}$_0$. 
Here we make use of Eq.(\ref{eqn019b}). 
In Eq.(\ref{eqn019b}) we notice that
for given $\alpha$ and $\beta$ either
${\cal F}_{\alpha\beta}$ (e.g. $\beta>0\rightarrow \alpha>0$) or 
${\cal G}_{\alpha\beta}$  (e.g. $\beta>0\rightarrow\alpha<0$) 
is not equal to zero but never both of them are zeros together. 
Then, we shall set (up to not essential 
normalization factor $1/\sqrt{2\omega_{\alpha {\bf k}_0}}$)
\begin{equation}
\begin{array}{l}
a_{11;1}=\cos ^2(\theta );a_{11;2}=\gamma _{+}^2\sin ^2(\theta
);b_{11;1}=0;b_{11;2}=-\gamma _{-}^2\sin ^2(\theta ), \\ 
a_{12;1}=a_{12;2}=-\gamma _{+}\sin (\theta )\cos (\theta
);b_{12;1}=b_{12;2}=0, \\ 
c_{11;1}=0;c_{11;2}=\gamma _{+}\gamma _{-}\sin ^2(\theta
);d_{11;1}=0;d_{11;2}=-\gamma _{+}\gamma _{-}\sin ^2(\theta ), \\ 
c_{12;1}=\gamma _{-}\sin (\theta )\cos (\theta
);c_{12;2}=0;d_{12;1}=0;d_{12;2}=-\gamma _{-}\sin (\theta )\cos (\theta ).
\end{array}
\end{equation}
Here we imply that $a_{\alpha\beta}$ and $b_{\alpha\beta}$ are nonzero
only for the particle-particle sector 
while  $c_{\alpha\beta}$ and $d_{\alpha\beta}$ are nonzero only
for the particle-antiparticle sector (i.e.
$c_{11}$ actually is $c_{1\rightarrow -1}$).

\begin{figure}
\centering
\begin{minipage}[c]{0.3\hsize}
\epsfig{width=\hsize,file=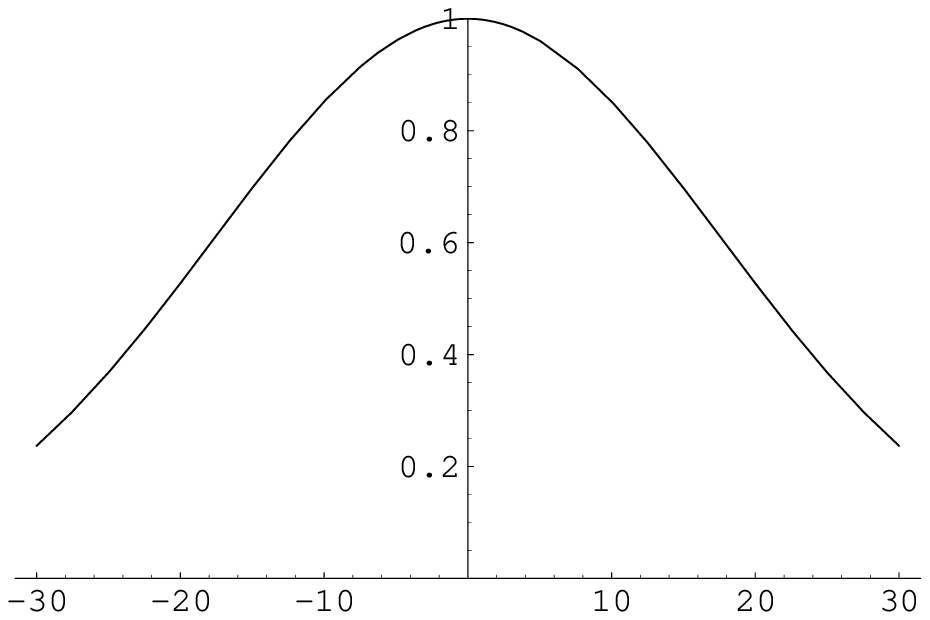}
\end{minipage}
\hspace*{0.5cm}
\begin{minipage}[c]{0.3\hsize}
\epsfig{width=\hsize,file=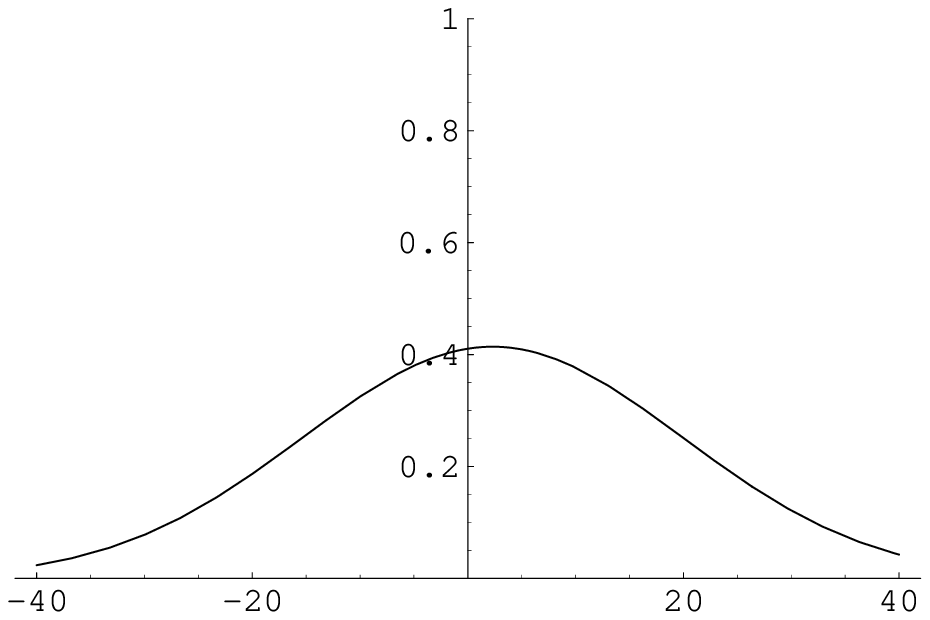}
\end{minipage}
\begin{minipage}[c]{0.3\hsize}
\epsfig{width=\hsize,file=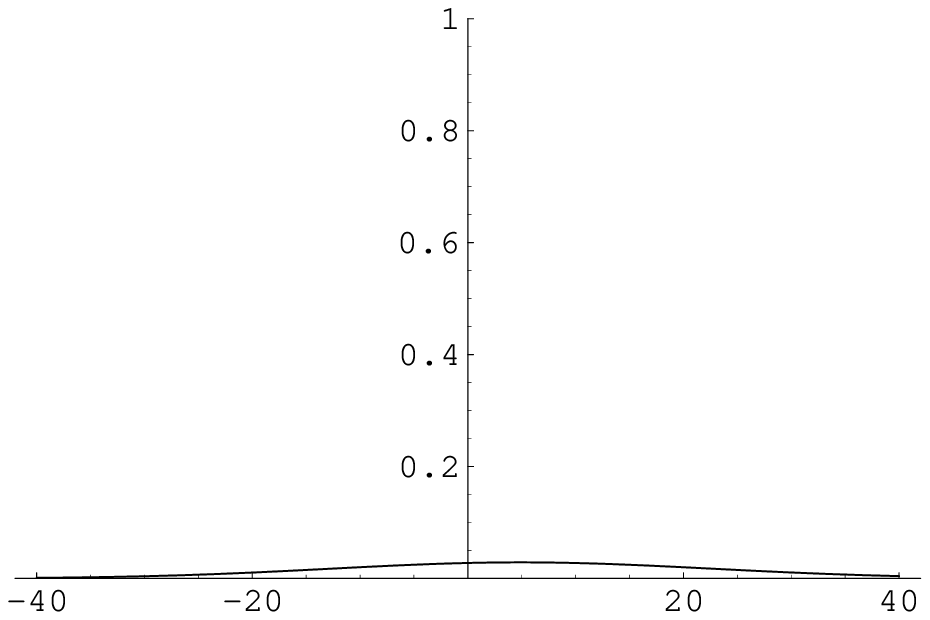}
\end{minipage}
\vspace*{0.5cm}
\begin{minipage}[c]{0.3\hsize}
\epsfig{width=\hsize,file=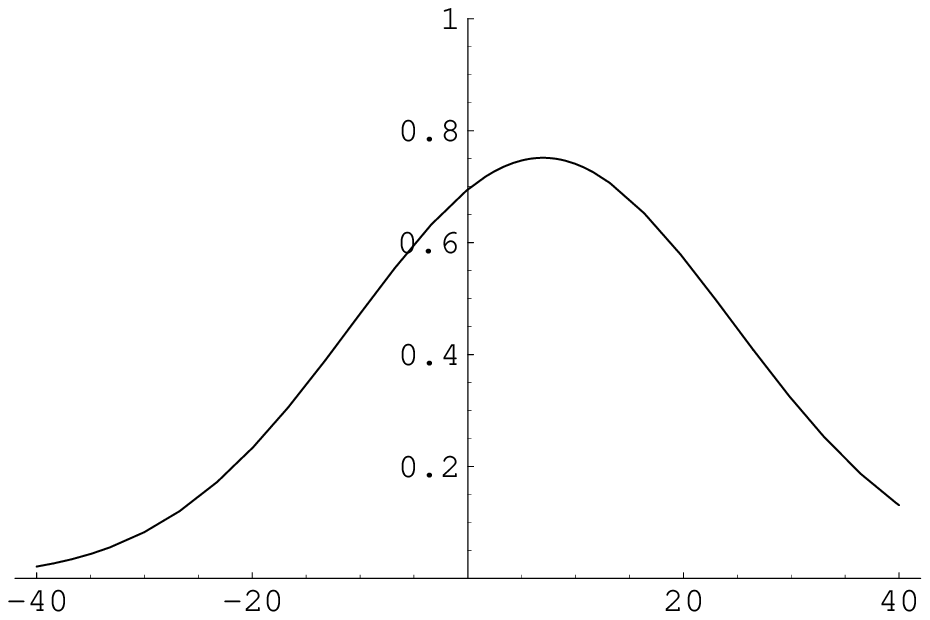}
\end{minipage}
\hspace*{0.5cm}
\begin{minipage}[c]{0.3\hsize}
\epsfig{width=\hsize,file=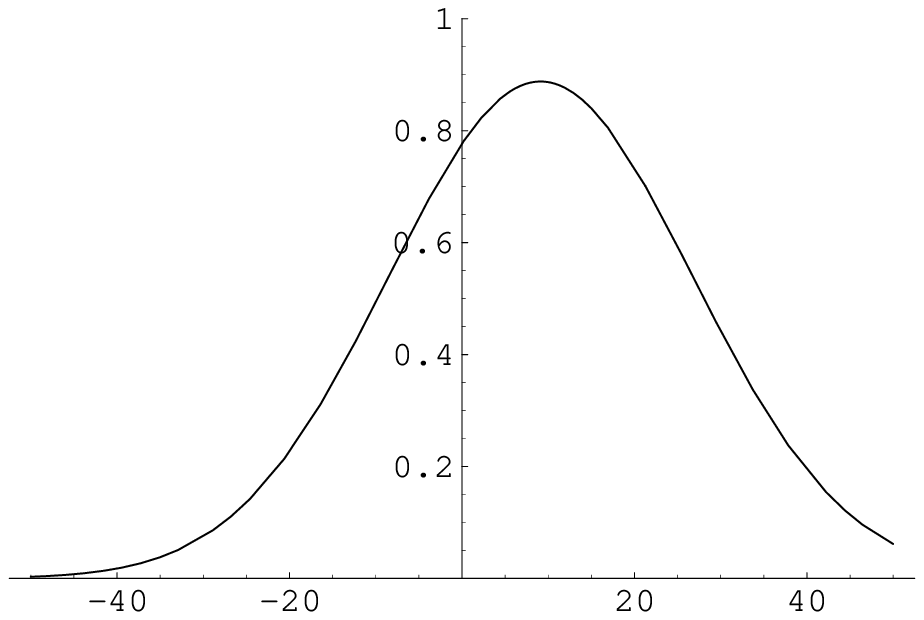}
\end{minipage}
\begin{minipage}[c]{0.3\hsize}
\epsfig{width=\hsize,file=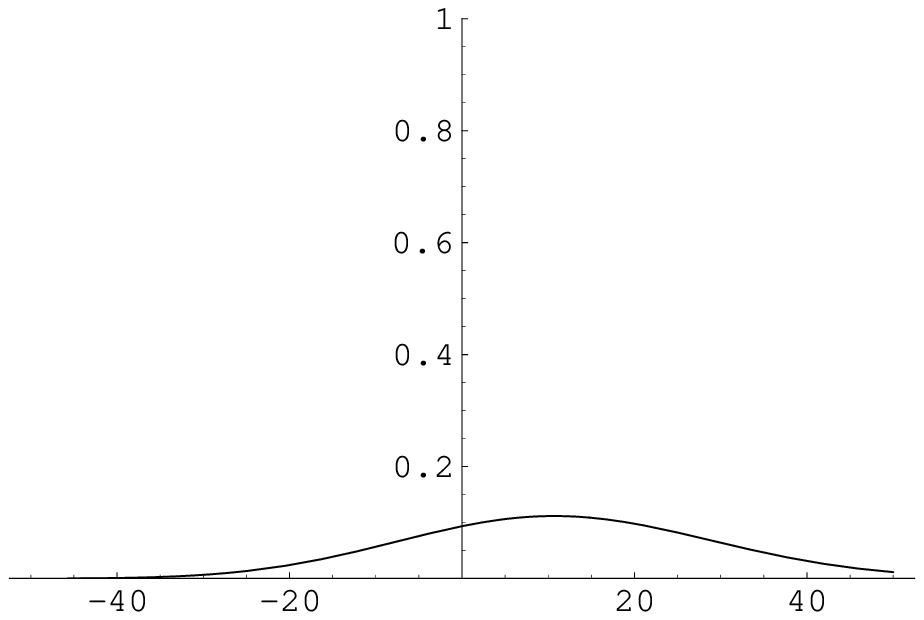}
\end{minipage}
\caption{
Example of propagation of gaussian wave-packet for particle of flavor 
$\alpha=1$ at $k_0\approx 0.35$GeV (time flow left-to-right and up-to-down).
Here and in what follows the distance scale is GeV$^{-1}$.
}
\label{plot04}
\end{figure}

\begin{figure}
\centering
\begin{minipage}[c]{0.3\hsize}
\epsfig{width=\hsize,file=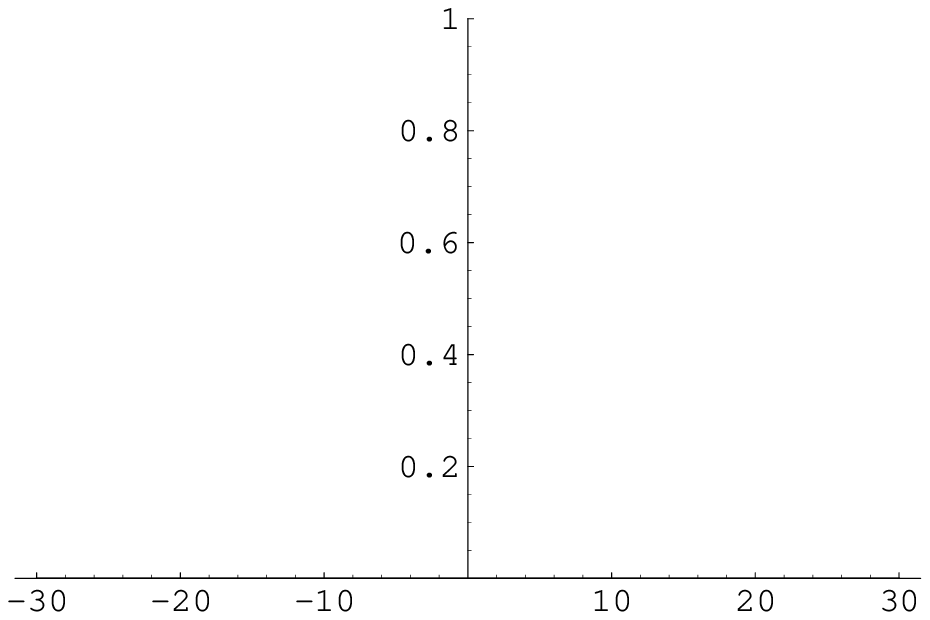}
\end{minipage}
\hspace*{0.5cm}
\begin{minipage}[c]{0.3\hsize}
\epsfig{width=\hsize,file=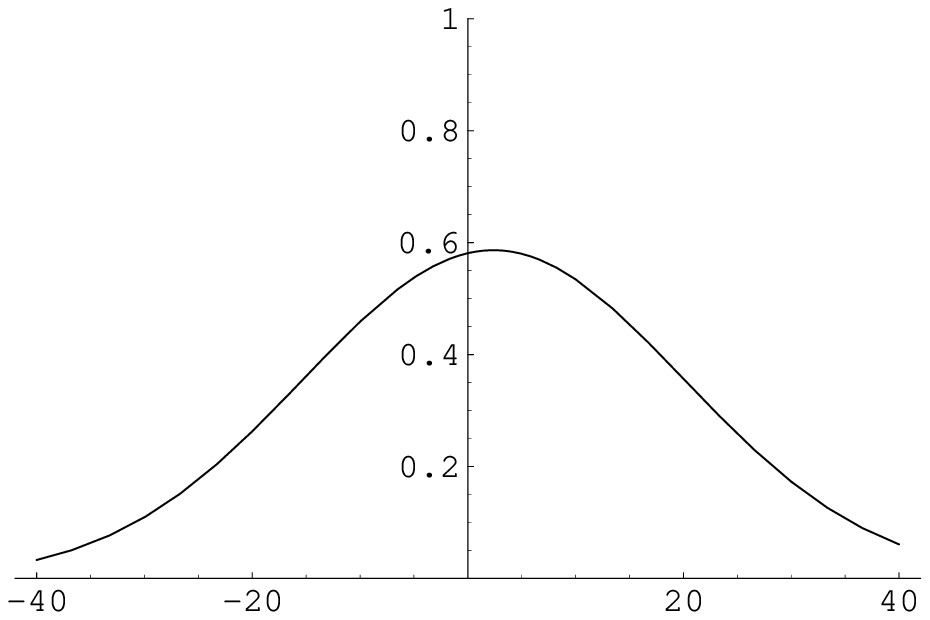}
\end{minipage}
\begin{minipage}[c]{0.3\hsize}
\epsfig{width=\hsize,file=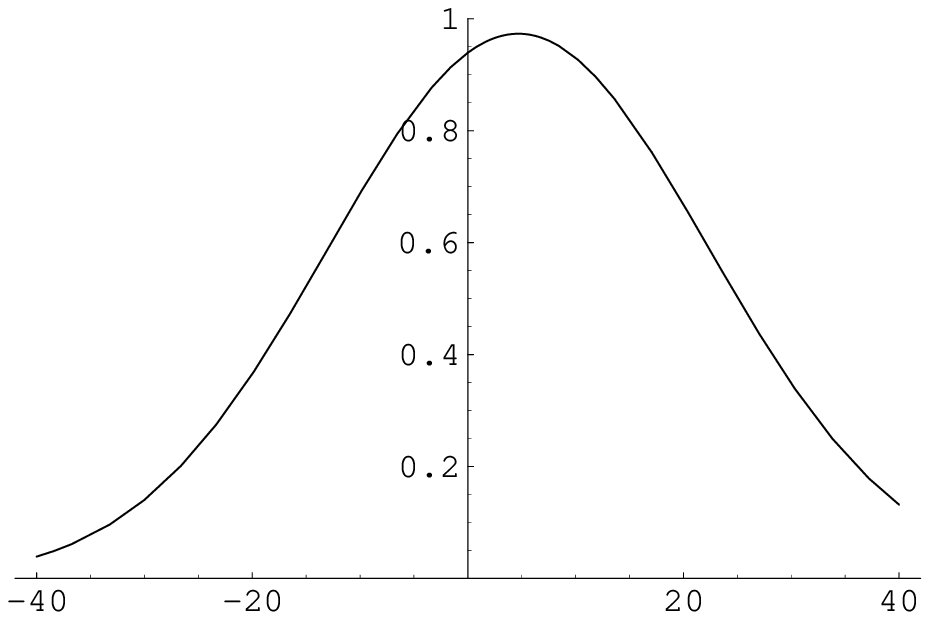}
\end{minipage}
\vspace*{0.5cm}
\begin{minipage}[c]{0.3\hsize}
\epsfig{width=\hsize,file=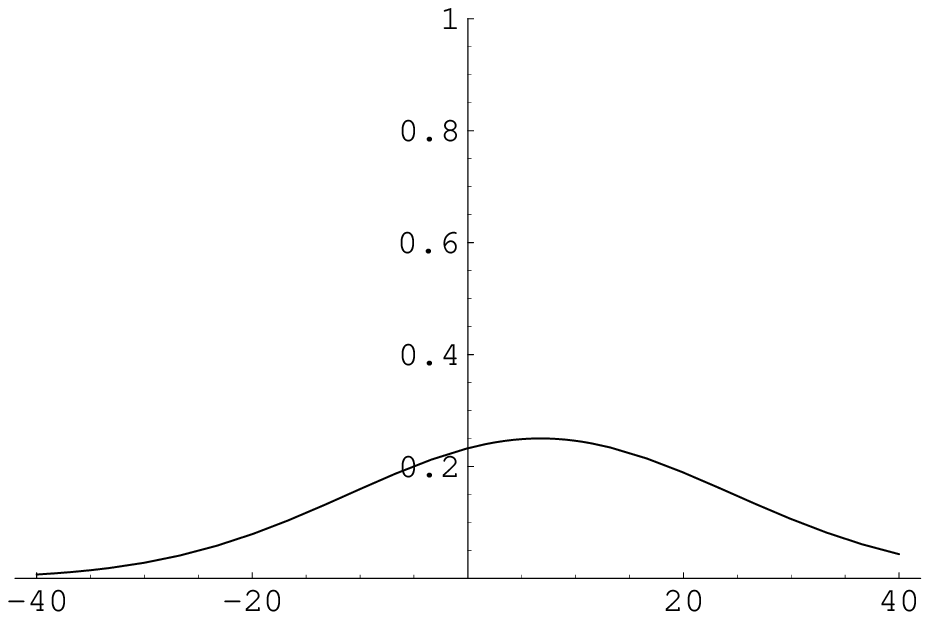}
\end{minipage}
\hspace*{0.5cm}
\begin{minipage}[c]{0.3\hsize}
\epsfig{width=\hsize,file=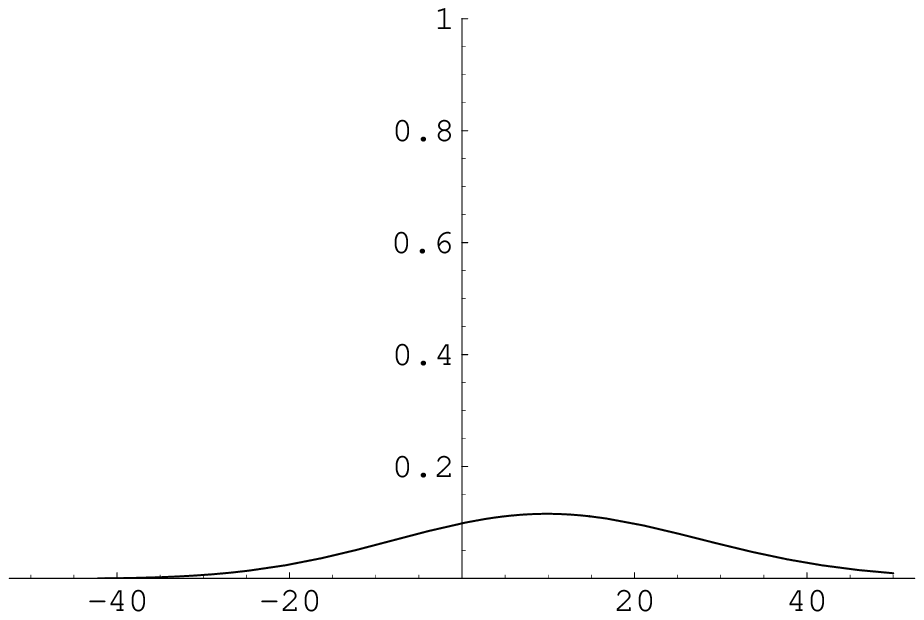}
\end{minipage}
\begin{minipage}[c]{0.3\hsize}
\epsfig{width=\hsize,file=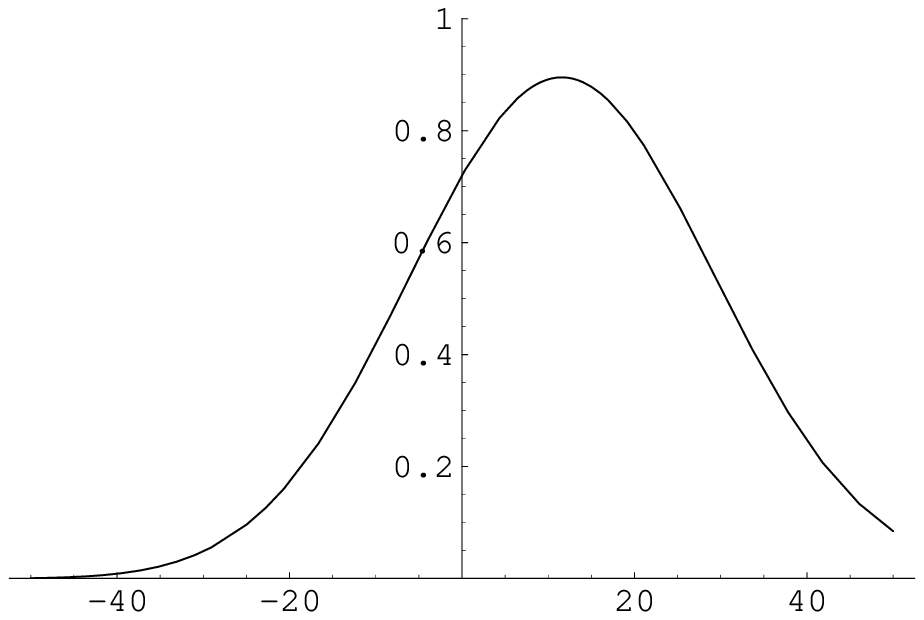}
\end{minipage}
\caption{Example of propagation of gaussian wave-packet
for particle of flavor 
$\alpha=2$ at $k_0\approx 0.35$GeV (time flow left-to-right and up-to-down).}
\label{plot05}
\end{figure}

\begin{figure}
\centering
\begin{minipage}[c]{0.42\hsize}
\epsfig{width=\hsize,file=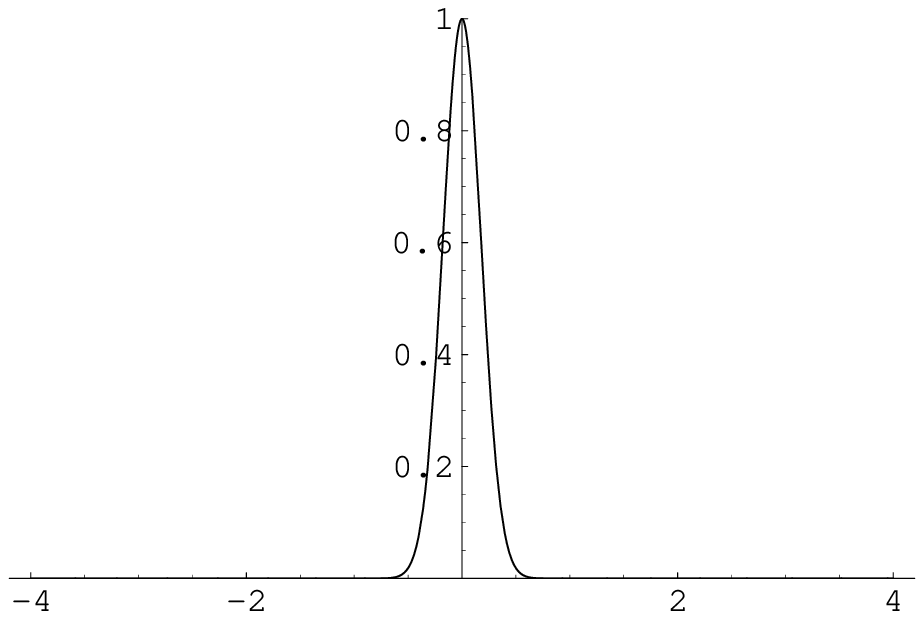}
\end{minipage}
\hspace*{0.5cm}
\begin{minipage}[c]{0.42\hsize}
\epsfig{width=\hsize,file=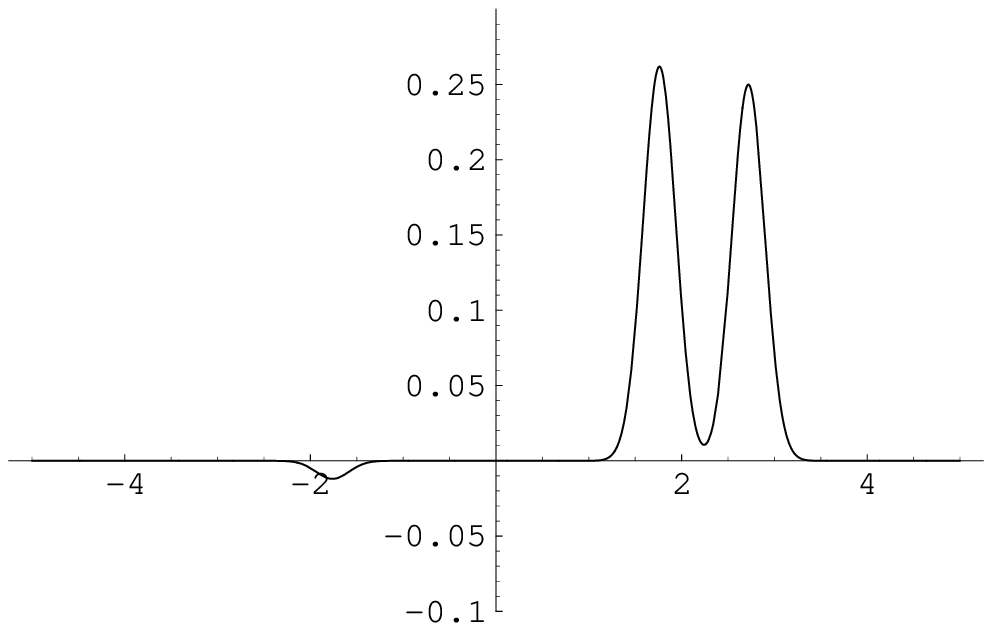}
\end{minipage}
\caption{
Two snapshots of distribution of flavor charge vs distance for
originally well-localized wave-packet give 
an example of the coherence loss by a point-like flavor source. 
Also, clearly seen is EPR-correlated
antiparticle wave-packet traveling in the opposite direction.
}
\label{plot01x}
\end{figure}

\begin{figure}
\centering
\begin{minipage}[c]{0.3\hsize}
\epsfig{width=\hsize,file=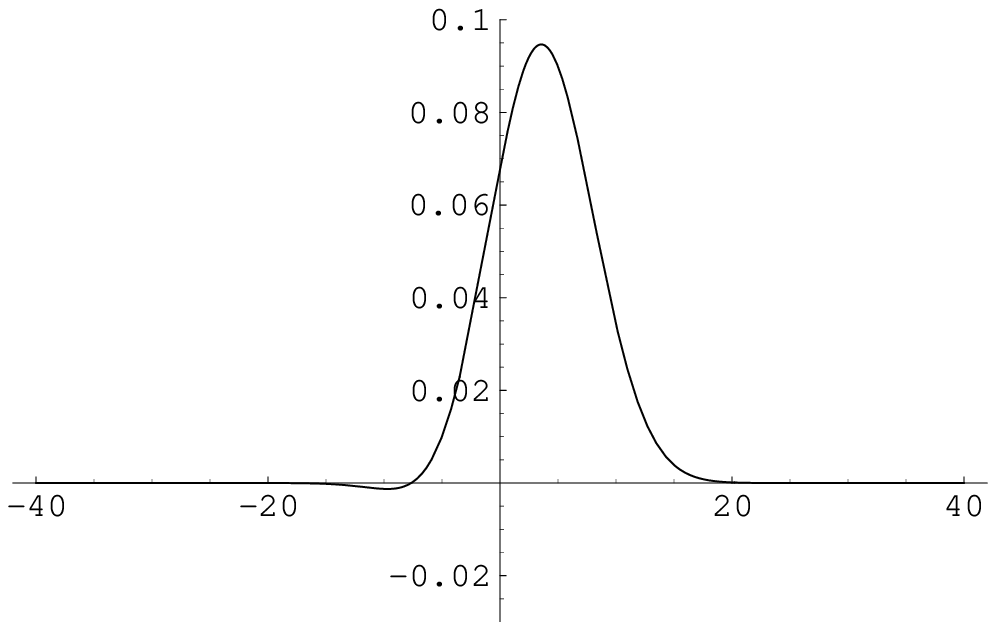}
\end{minipage}
\hspace*{0.5cm}
\begin{minipage}[c]{0.3\hsize}
\epsfig{width=\hsize,file=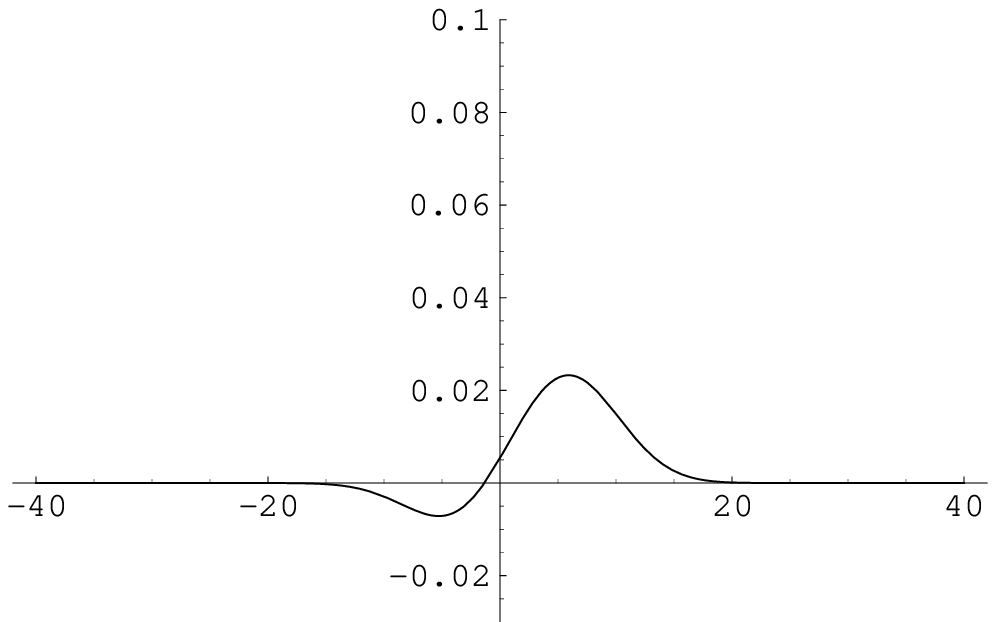}
\end{minipage}
\begin{minipage}[c]{0.3\hsize}
\epsfig{width=\hsize,file=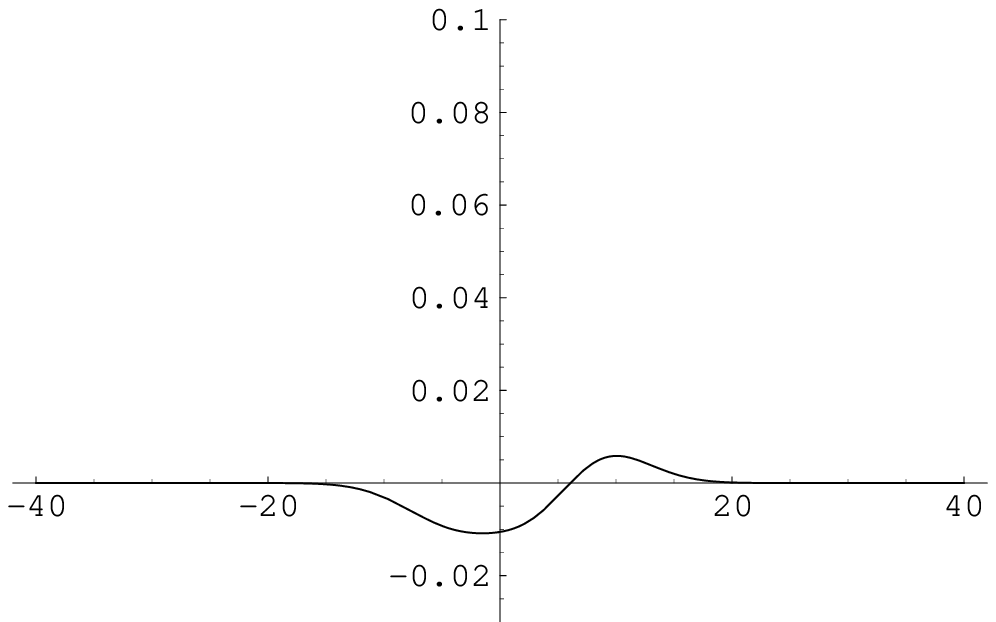}
\end{minipage}
\vspace*{0.5cm}
\begin{minipage}[c]{0.3\hsize}
\epsfig{width=\hsize,file=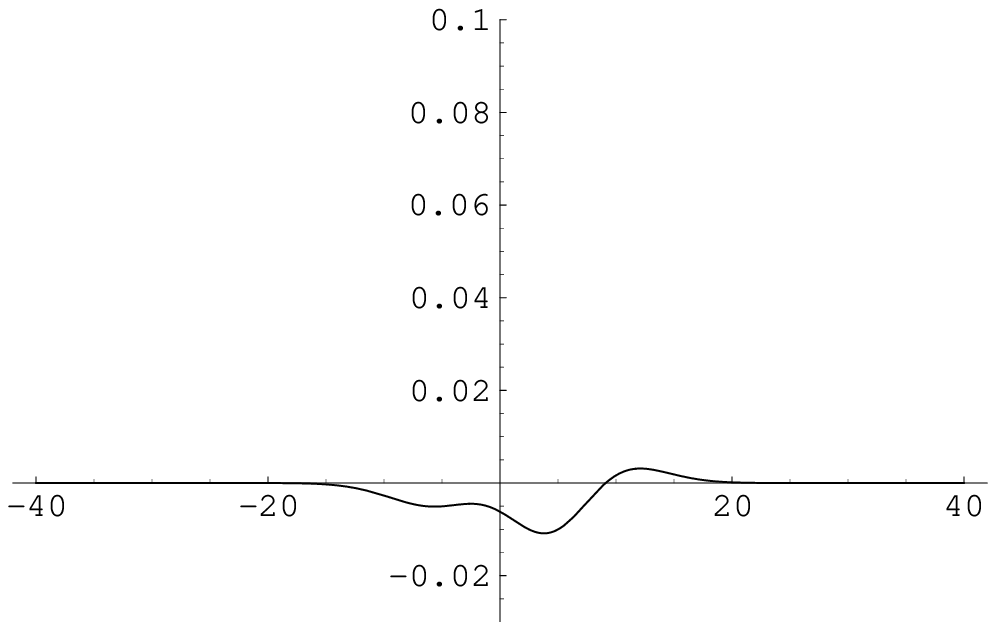}
\end{minipage}
\hspace*{0.5cm}
\begin{minipage}[c]{0.3\hsize}
\epsfig{width=\hsize,file=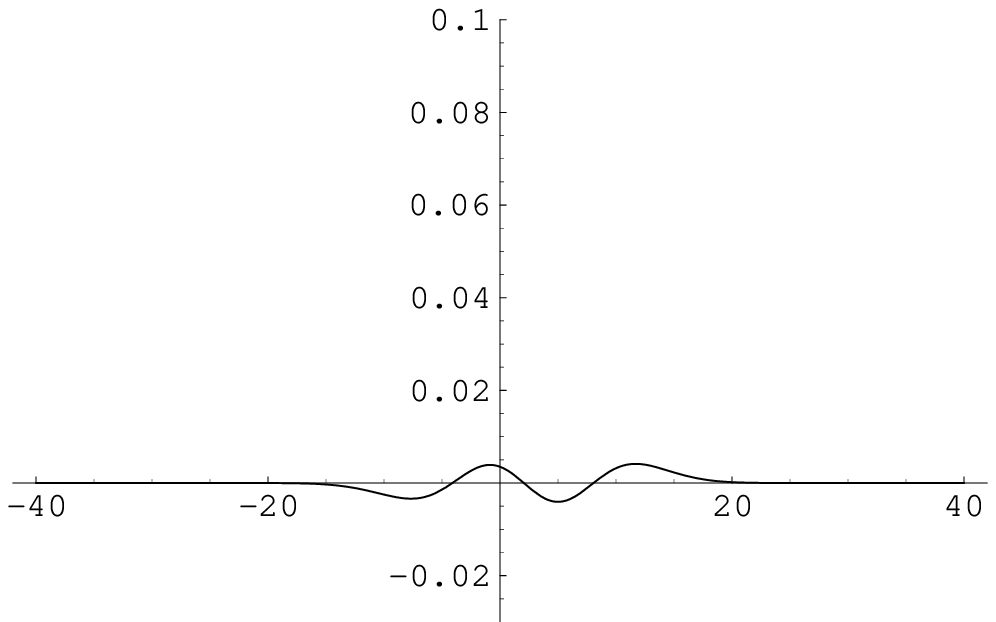}
\end{minipage}
\begin{minipage}[c]{0.3\hsize}
\epsfig{width=\hsize,file=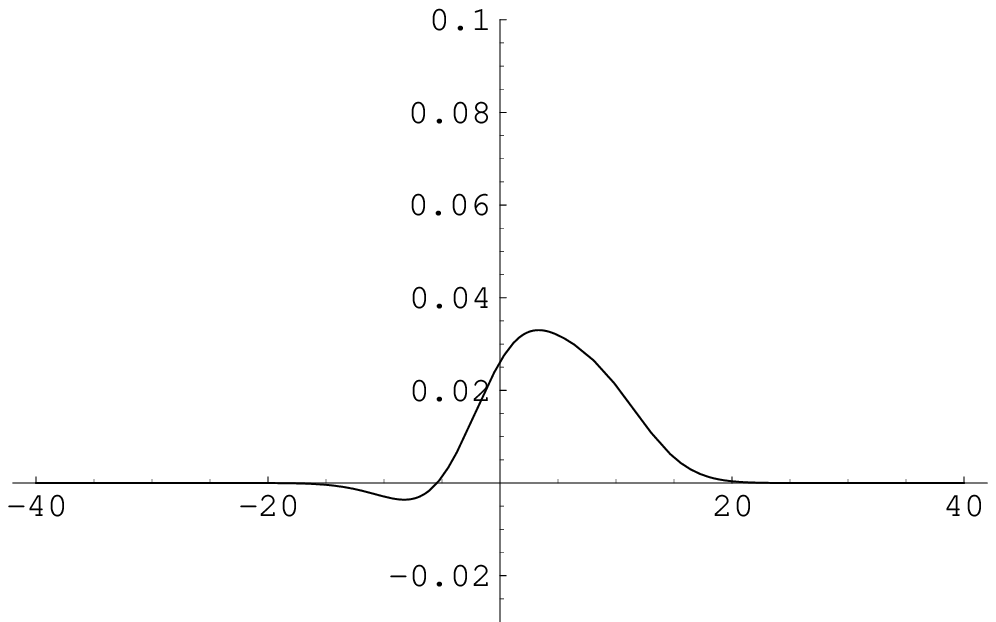}
\end{minipage}
\caption{Example of field-theoretical fluctuations at the "cross-over" 
time when particle of flavor $\alpha=1$ has been almost completely 
converted into flavor
$\alpha=2$ (time flow left-to-right and up-to-down).}
\label{plot06}
\end{figure}

To show the numerical results, we've chosen specific values of $m_1$ and 
$m_2$ close to the parameters of  
$\eta -\eta ^{\prime }$ system, i.e.
$m_1=540MeV$, $m_2=930MeV$ and $\theta =\pi /4$. We
simulated the  time evolution of the initial gaussian wave-packet 
according to Eq.(\ref{eqn019b}) for a range of incident particle energies. 
As can be seen in Figs.\ref{plot04} and \ref{plot05}, a typical wave-packet 
propagates in the right direction
oscillating at the same time into the other flavor. After a certain time, 
$\alpha=1$ particle is almost completely converted into 
$\alpha =2$ flavor after
which the reverse process takes place. 
With time evolution, the
original gaussian wave-packet deforms so that two separate
gaussians eventually emerge. This corresponds to two mass-eigenstates
completely separated in space: no flavor oscillations occur after
that point in agreement with the concept of coherence length. 
In particular, if $\sigma $ is large (or almost point source)
and $k_0\neq 0$,
the two mass-eigenstates separate almost immediately and 
propagate
independently producing no flavor oscillations at all [see Fig.\ref{plot01x}].
The additional effect due to the nontrivial flavor vacuum 
is rather small and is most noticeable only at the moments
when one of the flavors has almost completely disappeared due to 
the flavor conversion, as shown in Fig.\ref{plot06}. 
More remarkable
is the presence of the traces of negative flavor charge
propagating in the opposite direction to that of the main wave-packet.
This coherent beam of "recoil" anti-particles is due to the terms of the form 
$\exp[(-{\bf v}t-{\bf x})^2]$ in Eq.(\ref{eqn019b}); it is correlated with
the positive wing at all times via the mechanism similar to the EPR-effect, 
[see Fig.\ref{plot01x}]. 
The contribution from  
the high-frequency term, prominent in time-evolution, translates in space
as an interference between these parts of the wave-packet, propagating to 
the right and to the left respectively, and dies out almost immediately.

\begin{figure}
\centering
\epsfig{width=250pt,file=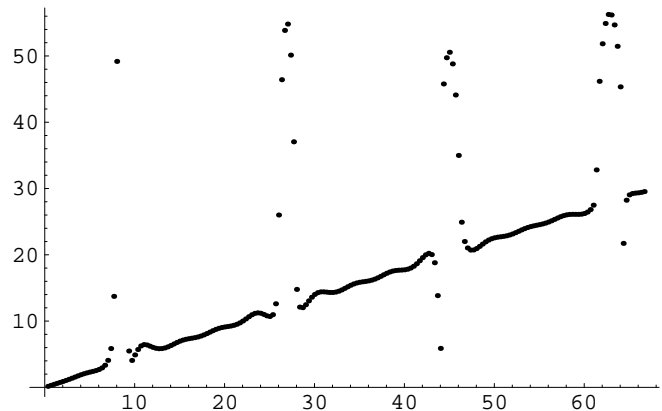}
\caption{An example of tracking
the position of maximum of the gaussian wave-packet.
Position (y-axis) vs time (x-axis) is plotted and disruptions
in numerical procedure
caused by field-theoretical fluctuations are visible.
}
\label{plotX}
\end{figure}

So far, we have studied the
propagation of a single flavor particle through space.
The space oscillations of flavor in conventional sense can be seen
only through the change in the amplitude of the wave-packet
as the particle flies through the space. To observe space
oscillations explicitly, we 
numerically traced with time
the position of the maximum of the wave-packet Eq.(\ref{eqn019b}). 
We found that the maximum propagates at
approximately constant speed consistent with 
\begin{equation}
\label{eqn028a}v=\frac{2k_0}{\sqrt{k_0^2+m_1^2}+\sqrt{k_0^2+m_2^2}}.
\end{equation}
In the numerical procedure we also have observed
interesting effect, namely, that at certain times the procedure
became unstable and produced significant fluctuations in
the position of the maximum, as shown in Fig.\ref{plotX}. 
After a closer examination
it turned out that this behavior was related to those
extremely short periods of time when one of the flavors 
almost completely disappeared so that the quantum field fluctuations 
became important and significantly distorted the shape of the
wave-packet and washed out the information about its maximum.
In Fig.\ref{plot06}, we presented a detailed look at this
short interval of time for $\alpha=1$ particle.
Other than that, the propagation of
the wave-packet is consistent with Eq.(\ref{eqn028a}). 

\begin{figure}
\centering
\begin{minipage}[c]{0.45\hsize}
\epsfig{width=\hsize,file=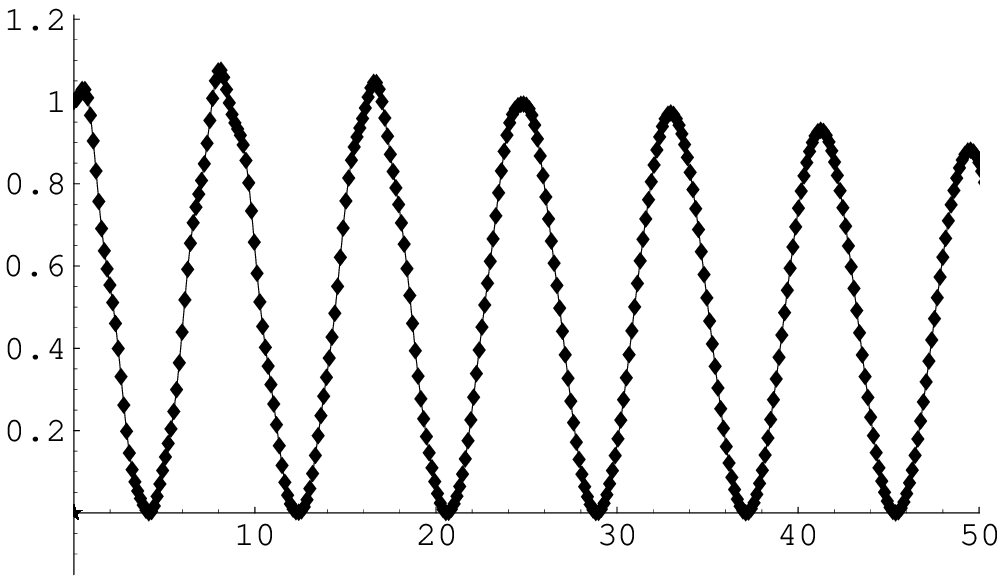}
\end{minipage}
\hspace*{0.5cm}
\begin{minipage}[c]{0.45\hsize}
\epsfig{width=\hsize,file=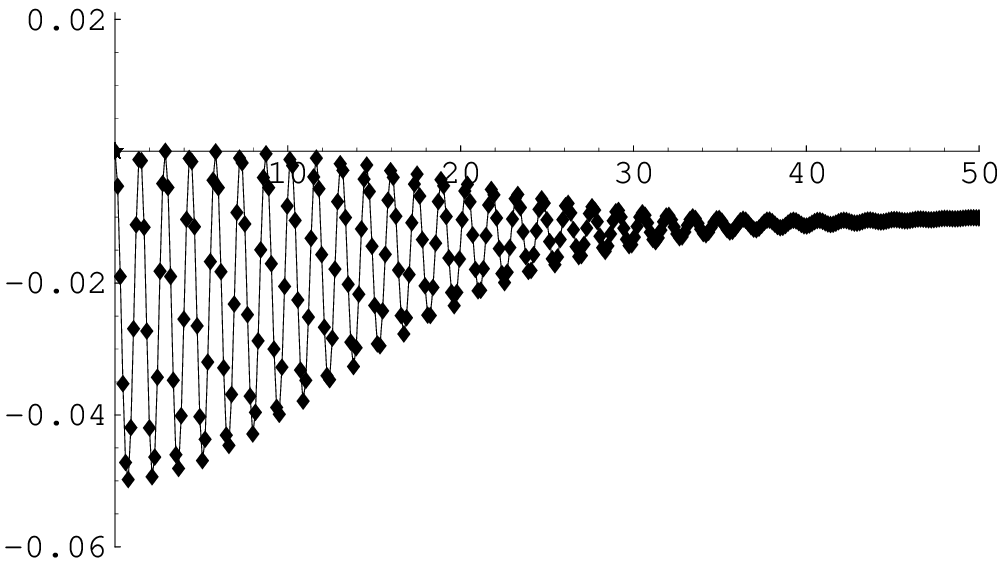}
\end{minipage}
\caption{
Space oscillations for particle (left) and 
antiparticle (right) populations
(for flavor $\alpha=1$). ($k\approx 0.35$GeV)}
\label{plot02b}
\end{figure}

\begin{figure}
\centering
\begin{minipage}[c]{0.45\hsize}
\epsfig{width=\hsize,file=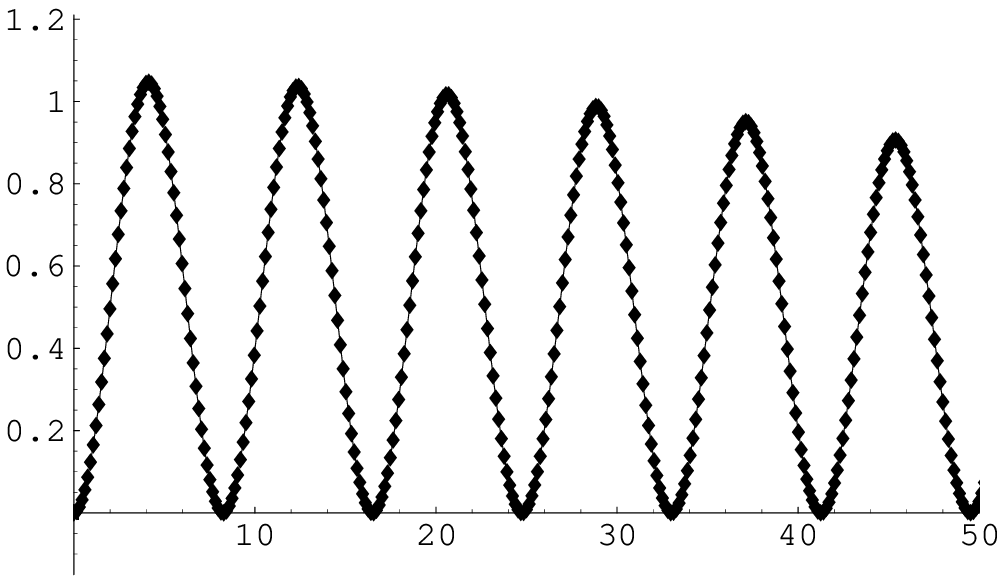}
\end{minipage}
\hspace*{0.5cm}
\begin{minipage}[c]{0.45\hsize}
\epsfig{width=\hsize,file=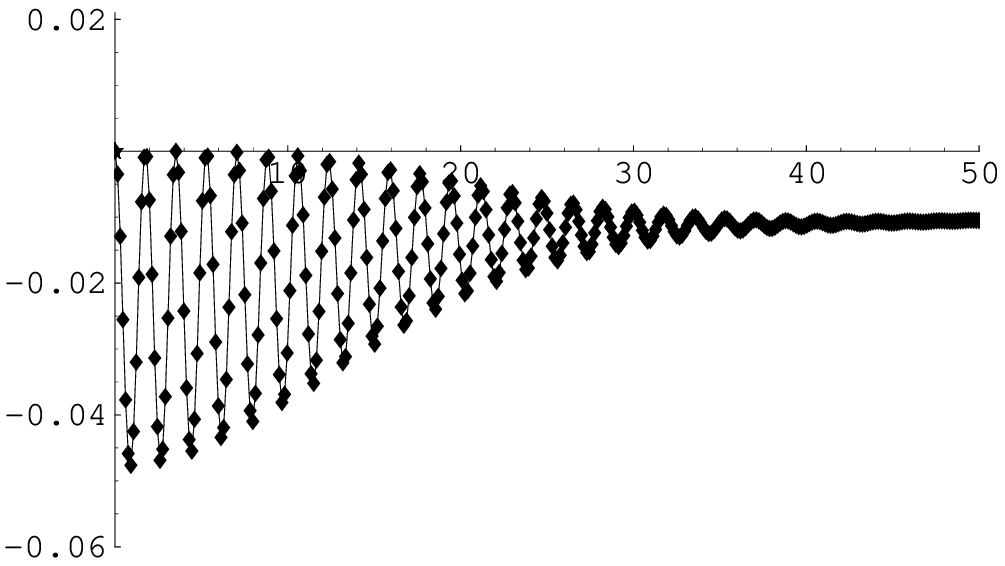}
\end{minipage}
\caption{
Space oscillations for particle (left) and 
antiparticle (right) populations
(for flavor $\alpha=2$). ($k\approx 0.35$GeV)}\label{plot03b}
\end{figure}

With that in mind, we reproduced the plot of the maximum amplitude of the
wave-packet vs. distance. 
In Figs.\ref{plot02b} and \ref{plot03b}, indeed,  we observe the space
oscillations with the wave-length $L\approx \frac{2k_0}{\Delta m^2}$. 
We also observed that, when the momentum of the particle is sufficiently
low, the form of the space-oscillations is noticeably
distorted from the quantum-mechanical prescription; 
for example, at certain points the flavor
charge becomes negative. However, although in the time-dynamics
the nontrivial vacuum effect introduces significant corrections,
the quantum-mechanical formula generally work fine 
for the space-oscillation and the field-theoretical 
corrections are less noticeable.
Also, the field-theoretical corrections decrease as the energy increases and
die out with the distance.

\section{Conclusion}

We studied the problem of time-to-space conversion of field-theoretical flavor
oscillation formula with nontrivial vacuum effect. 
We approached this problem from the most general
wave-packet approach utilizing the canonical formalism of
creation/annihilation operators. This allowed us to account for the 
nontrivial flavor vacuum effect,
which is otherwise lost. We derived the space-oscillation formula for
a flavor particle initially created as a sharp wave-packet.
Different from time-oscillations, where 
the flavor vacuum effect introduces a 
prominent high-frequency term, 
 we found no major differences 
from quantum mechanical prescription
in the case of space oscillations in field theory. Unlike in
time-dimension, only single
mode is found in space-dimensions with the
wave-length consistent with the quantum mechanics. 
Certain quantitative and qualitative differences,
however, are present in the shape 
and  especially  in the antiparticle content of the space oscillations
in quantum field theory.

We further applied our general formalism to a specific case of 
mixing of two bosons,
in which we considered in great details the evolution of the flavor-particle
wave-packet. We observed the space 
oscillations of flavor
through the dependence of maximum on
distance. We found that the propagation of the flavor particle
in the above setting is
consistent with the group velocity $v\approx k/E$. 
We observed small differences between quantum-mechanical
and field-theoretical results for flavor oscillations and found that 
the flavor charge may
become negative at certain points in space. Also, a correlated
beam of antiparticles, propagating in the direction opposite to that of 
the main wave-packet, was noticed in our numerical simulation. 

\acknowledgments

This work was supported in part by a grant from the U.S. Department of 
Energy(DE-FG02-96ER 40947). Y.M. was also supported in part by the 
SURA/JLAB Fellowship.


\newpage

\end{document}